\documentclass[11pt]{article}

\usepackage[utf8]{inputenc}
\usepackage[T1]{fontenc}
\usepackage{textcomp}
\usepackage{lmodern}
\usepackage[nopatch=footnote,babel=false]{microtype}

\DeclareUnicodeCharacter{2264}{\ensuremath{\leq}}   
\DeclareUnicodeCharacter{2265}{\ensuremath{\geq}}   
\DeclareUnicodeCharacter{223C}{\ensuremath{\sim}}   
\DeclareUnicodeCharacter{0301}{\'{}}                 

\usepackage[margin=1in]{geometry}
\usepackage{graphicx}
\usepackage{amsmath,amssymb}
\usepackage{booktabs}
\usepackage{multirow}
\usepackage{float}
\usepackage{hyperref}
\usepackage{caption}
\usepackage{subcaption}
\usepackage{authblk}
\usepackage{setspace}
\usepackage{longtable}
\usepackage{array}

\hypersetup{
    colorlinks=true,
    linkcolor=blue,
    citecolor=blue,
    urlcolor=blue
}

\newcommand{\figph}[4][ht]{%
  \begin{figure}[#1]
    \centering
    \IfFileExists{#2}{%
      \includegraphics[width=0.9\textwidth]{#2}%
    }{%
      \fbox{\begin{minipage}[c][0.30\textheight][c]{0.85\textwidth}
        \centering\small
        Placeholder for figure file: \texttt{#2}\\[6pt]
        (Replace with the actual image file)
      \end{minipage}}%
    }%
    \caption{#3}%
    \ifx\relax#4\relax\else\label{#4}\fi
  \end{figure}
}

\title{\textbf{Pan-Cancer Mapping of the Tumor Immune Landscape through Metagene Clustering and Predictive Modeling}}
\author[]{Soham Chatterjee}
\date{August 2025}

\begin{document}
\maketitle

\begin{abstract}
As immunotherapies become standard cancer treatments, it is increas-
ingly important to identify a patient's immune profile, which encompasses
the activity of immune cells within the tumor microenvironment and the
presence of specific biomarkers. However, we lack mechanistic explana-
tions of the factors driving immune phenotypes. Despite advances in im-
mune profiling with high-throughput sequencing, the mechanisms driving
specific immune phenotypes remain unclear.

This study aimed to identify novel, robust immune-related gene clus-
ters (metagenes) and evaluate their prognostic significance and functional
relevance across various pan-cancer types using a comprehensive com-
putational pipeline. We acquired pan-cancer bulk RNA-seq and estab-
lished immune subtypes from The Cancer Genome Atlas (TCGA). Using
expression-based filtering and clustering of genes with ANOVA and Gaus-
sian Mixture Model (GMM), we identified 48 unique metagenes. These
metagenes achieved an 87\% accuracy in predicting the established sub-
types.

SHAP analysis revealed the most predictive metagenes per subtype,
while functional enrichment analysis identified their associated pathways.
Additionally, genes were ranked by differential expression between high-
and low-expression groups. The metagenes revealed insights, including the
co-expression of immune activation and regulatory factors, links between
cell cycle regulation and immune evasion, and dynamic microenvironment
remodeling signatures. Kaplan-Meier survival analysis and multivariate
Cox Regression revealed that many metagenes had prognostic value for
overall survival.

Overall, the metagenes represent coordinated biological programs across
diverse cancer types, providing a foundation for developing robust, broadly
applicable immuno-oncology biomarkers that extend beyond single-gene
markers. They demonstrate prognostic value across cancer types and hold
potential to guide immunotherapy treatment decisions.
\end{abstract}

\section{Introduction}

The tumor immune microenvironment is a heterogeneous, evolving ecosystem
of immune cells and their interactions with cancer cells and associated tumor
components. Cancer progression is heavily dependent on the balance between
pro- and anti-tumor immune responses \cite{anderson2020,ostrand2008,johansson2008}. Characterizing a patient's im-
mune state has a significant impact on clinical outcomes, particularly as im-
munotherapies become standard treatments for multiple cancer types \cite{gnjatic2017,esfahani2020}.
However, current tumor immune state classifications interpret descriptive cat-
egories (immune-hot, immune-cold, immune-excluded) that provide limited in-
sight into the biological programs driving these phenotypes \cite{chen2017,mellman2023}. This limita-
tion is particularly problematic, as immunotherapy response rates remain low
(15--30\%) in most cancers despite growing success \cite{das2019}. Moreover, the lack of
mechanistic pan-cancer insights limits the development of broadly applicable
therapeutic strategies.

High-throughput technologies such as RNA-sequencing enable comprehen-
sive profiling of the tumor microenvironment, offering detailed insights into
tumor-immune interactions \cite{hong2020}. Current analytical approaches often involve
single-gene biomarkers, such as PD-1 expression or CD8+ T-cell counts. How-
ever, these may fail to capture the multiple pathways and factors determining
immunotherapy response, which helps explain why immunotherapy response
rates are poorly predicted \cite{arora2019,pilard2021}. Using existing immune-related gene sets, on
the other hand, may overlook broader, coordinated responses not covered by
our current understanding of tumor biology.

This study addresses these limitations through a data-driven framework for
the identification of transcriptional programs characterizing pan-cancer immune
states. We leverage The Cancer Genome Atlas with the work of Thorsson et al. \cite{thorsson2019}, which characterized six broad pan-cancer immune subtypes. Inte-
grating this data with advanced machine learning and bioinformatics techniques,
we aimed to identify functionally distinct gene clusters (metagenes) that pro-
vide mechanistic explanations of the different immune subtypes across many
cancers, while also validating their prognostic power and identifying key driver
genes. Through this framework, we identified 48 unique metagenes that rep-
resent coordinated biological programs, able to characterize tumor phenotypes.
The metagenes achieved an 87\% accuracy in classifying the immune subtypes, as
well as revealed novel insights, including the co-expression of immune activation
and regulatory signals, the integration of tumor-intrinsic regulators with im-
mune evasion, and dynamic microenvironment remodeling networks. Through
this study, we aim to uncover coordinated biological programs that can char-
acterize the tumor immune microenvironment across multiple cancer types, as
well as have prognostic relevance that makes them suitable for future biomarker
discovery.

\section{Materials and Methods}

\subsection{Pan-Cancer Data Acquisition and Preprocessing}

Three pan-cancer datasets were acquired from The Cancer Genome Atlas through
the UCSC Xena Browser (pancanatlas.xenahubs.net):

\begin{enumerate}
    \item Batch effects normalized mRNA dataset: This dataset contains bulk
    RNA-seq data for 20,531 genes across 11,060 patient samples, with units
    in \(\log_2(\)normalized counts + 1\()\).
    \item Immune subtype dataset: This dataset maps 9,126 patient samples
    (all included in the RNA-seq dataset) to their immune subtype (C1--C6).
    These subtypes were developed and assigned to each patient by Thors-
    son et al. \cite{thorsson2019}. Table \ref{tab:subtype_distribution} shows the distribution of patient samples across subtypes.
    \item Curated clinical dataset: This dataset includes clinical data for 12,591
    patient samples, 9,104 of which are included in the Immune subtype
    dataset. It provides multiple valuable clinical variables, such as over-
    all survival, progression-free survival, tumor type and stage, gender, and
    race.
\end{enumerate}

The RNA-seq dataset was filtered to keep only data for the 9,126 patient samples
with predefined immune subtypes. Data preprocessing involved KNN imputa-
tion (k=5) and removal of low-variance genes (bottom 40\%), yielding 12,320
genes.

\begin{table}[H]
\centering
\caption{Distribution of patient samples across subtypes}
\label{tab:subtype_distribution}
\begin{tabular}{lr}
\toprule
Subtype & Number of Patient Samples \\
\midrule
C1 & 2416 \\
C2 & 2591 \\
C3 & 2397 \\
C4 & 1157 \\
C5 & 385 \\
C6 & 180 \\
\midrule
Total & 9126 \\
\bottomrule
\end{tabular}
\end{table}

\subsection{Gene Clustering and Metagene Identification}

\paragraph{Feature Reduction}
To identify genes whose expression differed significantly across immune subtypes, a one-way Analysis of Variance (ANOVA) test was performed for each gene. ANOVA measures whether the mean expression of the genes between different clusters varies more than would be expected by chance. Adjusted p-values were calculated with Benjamini-Hochberg correction to control for multiple hypothesis testing.

Nearly all the genes (12,317 out of 12,320) had an adjusted p-value less than 0.05. As a result, a stricter cutoff was chosen for significant genes. 4,489 genes had an adjusted p-value of 0.0. These genes were selected for further analysis. The transposed gene expression matrix contained gene expression values for these 4,489 genes across the 9,126 patients.

\paragraph{Gene Clustering}
PCA was applied to reduce dimensionality from 4,489 genes to 50 principal components, preserving major expression patterns. Gaussian Mixture Model (GMM) clustering of these components produced 150 co-expressed gene clusters (metagenes). The full list of genes comprising each metagene is provided in the supplementary materials.

\subsection{Predictive Model Development}

\paragraph{Signature Score Calculation}
Three signature scores (mean, median, and standard deviation) were calculated for each metagene per patient, creating a 9,126 × 450 feature matrix.

\paragraph{ML Model Selection}
We selected three complementary ML models for immune subtype predictions: SVM (high-dimensionality handling), Random Forest (ensemble stability), and XGBoost (feature interactions).

\paragraph{Feature Selection}
We employed Recursive Feature Elimination with Cross Validation (RFECV) to select the optimal feature subset. RFECV iteratively removes features one by one and evaluates model performance to find the optimal feature set.

\paragraph{Hyperparameter Optimization}
To find the optimal hyperparameters for the ML models, hyperparameter tuning was performed using Optuna. Table \ref{tab:model_summary} summarizes the machine learning models, the RFECV-selected features, the number of unique metagenes among selected features, and the optimized hyperparameters.

\begin{table}[H]
\centering
\caption{Summary of machine learning models, including RFECV-selected features, number of unique metagenes among selected features, and hyperparameters optimized through tuning.}
\label{tab:model_summary}
\begin{tabular}{llllr}
\toprule
Model & Features & Metagenes & Hyperparameter & Value \\
\midrule
\multirow{2}{*}{SVM} & \multirow{2}{*}{46} & \multirow{2}{*}{36} & C & 2.2165 \\
 &  &  & Degree & 4 \\
\midrule
\multirow{2}{*}{Random Forest} & \multirow{2}{*}{56} & \multirow{2}{*}{38} & n\_estimators & 188 \\
 &  &  & Max depth & 10 \\
\midrule
\multirow{4}{*}{XGBoost} & \multirow{4}{*}{68} & \multirow{4}{*}{48} & n\_estimators & 331 \\
 &  &  & Max depth & 5 \\
 &  &  & Learning rate & 0.1547 \\
\bottomrule
\end{tabular}
\end{table}

\paragraph{Model Training and Evaluation}
Each model was trained with its optimal feature subset and tuned hyperparameters and evaluated using K-fold cross-validation (k=5). Multiple metrics were used to evaluate model performance. These include accuracy, precision, recall, F1 score, and ROC AUC. For all metrics, both training and test metrics were calculated. Additionally, for all metrics except accuracy, both an overall score and a macro score were calculated. These metrics were calculated per fold of data and were averaged.

\subsection{Functional Analysis: SHAP}

Shapley Additive Explanations (SHAP) was used to quantify the predictive value of each feature. SHAP calculates how much each feature contributes to each individual prediction, allowing interpretation of which metagenes drive classification decisions for each sample.

For the best-performing model, SHAP Violin plots were created for each cluster. These plots showed which features contributed most to the predictions for each immune subtype.

\subsection{Survival Analysis}

Kaplan-Meier survival analysis and Cox Proportional Hazards Regression were employed to assess the prognostic value of the metagenes.

\subsubsection{Kaplan-Meier Survival Analysis}
We used Kaplan-Meier analysis to investigate associations between metagene expression and survival. From the clinical dataset, 9,104 patient samples overlapping with the immune subtype dataset were obtained with overall survival status (OS) and time (OS.time, days); 9,074 had both defined.

A matrix of the 9,074 patient samples with their overall survival status and time (right-censored at death or last follow-up) was constructed. For each of the 68 features from XGBoost, samples were separated into High and Low groups by the median feature score. KM survival curves were estimated for each feature, and the 5-year survival probability was extracted for each group.

Log-rank tests with BH correction were conducted to assess whether survival differences between the two groups over time were statistically significant. KM plots were generated for each feature, and the statistical information (p-values, adjusted p-values, and survival probabilities) was saved for further analysis.

\subsubsection{Cox Proportional Hazards Regression}
We used multivariate Cox Regression to assess the prognostic value of the metagenes while adjusting for clinical confounders. The covariates included the 68 metagene-derived features along with gender, age at diagnosis (continuous), and cancer type (categorical; TCGA abbreviation) as clinical covariates to adjust for confounding.

Overall survival status and time were extracted from the clinical dataset. Samples with any missing values in survival data or clinical covariates were dropped, leaving 9055 out of the 9126 samples to perform Cox regression.

Hazard ratios (HRs), 95\% confidence intervals, and p-values were extracted from each covariate. The p-values were adjusted for multiple hypothesis testing using BH correction, and covariates with adjusted p-values \(\leq\) 0.05 were considered statistically significant.

\subsection{Functional Enrichment Analysis}

To identify the biological functions and associated pathways of the metagenes, gene ontology and pathway enrichment analysis was conducted on the predictive metagenes derived from SHAP.

All metagene-based SHAP features were compiled and mapped to unique metagene clusters, with each metagene represented once regardless of duplicates or summary statistics.

We extracted the gene set for each metagene and performed overrepresentation analysis using g:Profiler (biit.cs.ut.ee/gprofiler/gost) to identify statistically overrepresented GO terms and biological pathways (e.g., from KEGG, Reactome) within a gene set.

\subsection{Gene-Level Analysis Within Metagenes}

Gene-level analysis was performed on the 48 unique metagenes from XGBoost to identify the individual genes driving each metagene's biological signal. For each metagene, a representative score was selected per sample: mean if available, otherwise median, and if neither was present, standard deviation. The 9,126 samples were stratified into high and low groups based on the top and bottom quartiles (75th and 25th percentiles, respectively) of the metagene score distribution.

The mean expression of the high and low groups was calculated for each gene within the metagene. The difference in mean expression of each gene between the two groups was used to assess the relative upregulation or downregulation of each gene. The genes were ranked by the magnitude of this differential expression for each metagene. This allowed us to rank the genes most strongly driving the signal of each gene expression program.

\section{Results}

\subsection{Metagene Identification and Characterization}

\begin{figure}[ht]
  \centering
  \begin{subfigure}[t]{0.47\textwidth}
    \centering
    \IfFileExists{tsne_anova_genes.png}{%
      \includegraphics[width=\textwidth]{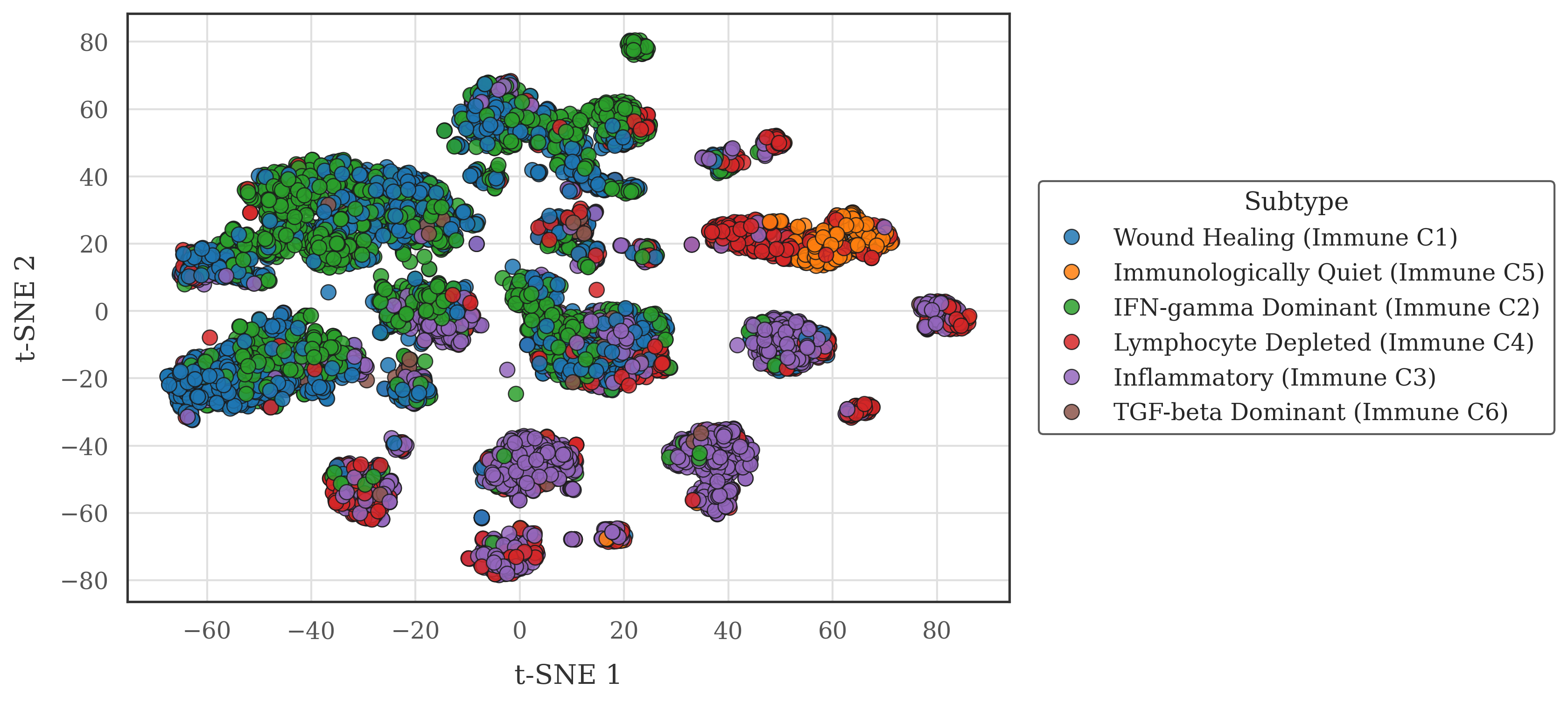}%
    }{%
      \fbox{\begin{minipage}[c][0.28\textheight][c]{\textwidth}
        \centering\small
        Placeholder:\\[4pt]\texttt{tsne\_anova\_genes.png}
      \end{minipage}}%
    }
    \caption{t-SNE visualization of ANOVA-selected genes.}
    \label{fig:tsne_anova}
  \end{subfigure}
  \hfill
  \begin{subfigure}[t]{0.47\textwidth}
    \centering
    \IfFileExists{tsne_xgboost_genes.png}{%
      \includegraphics[width=\textwidth]{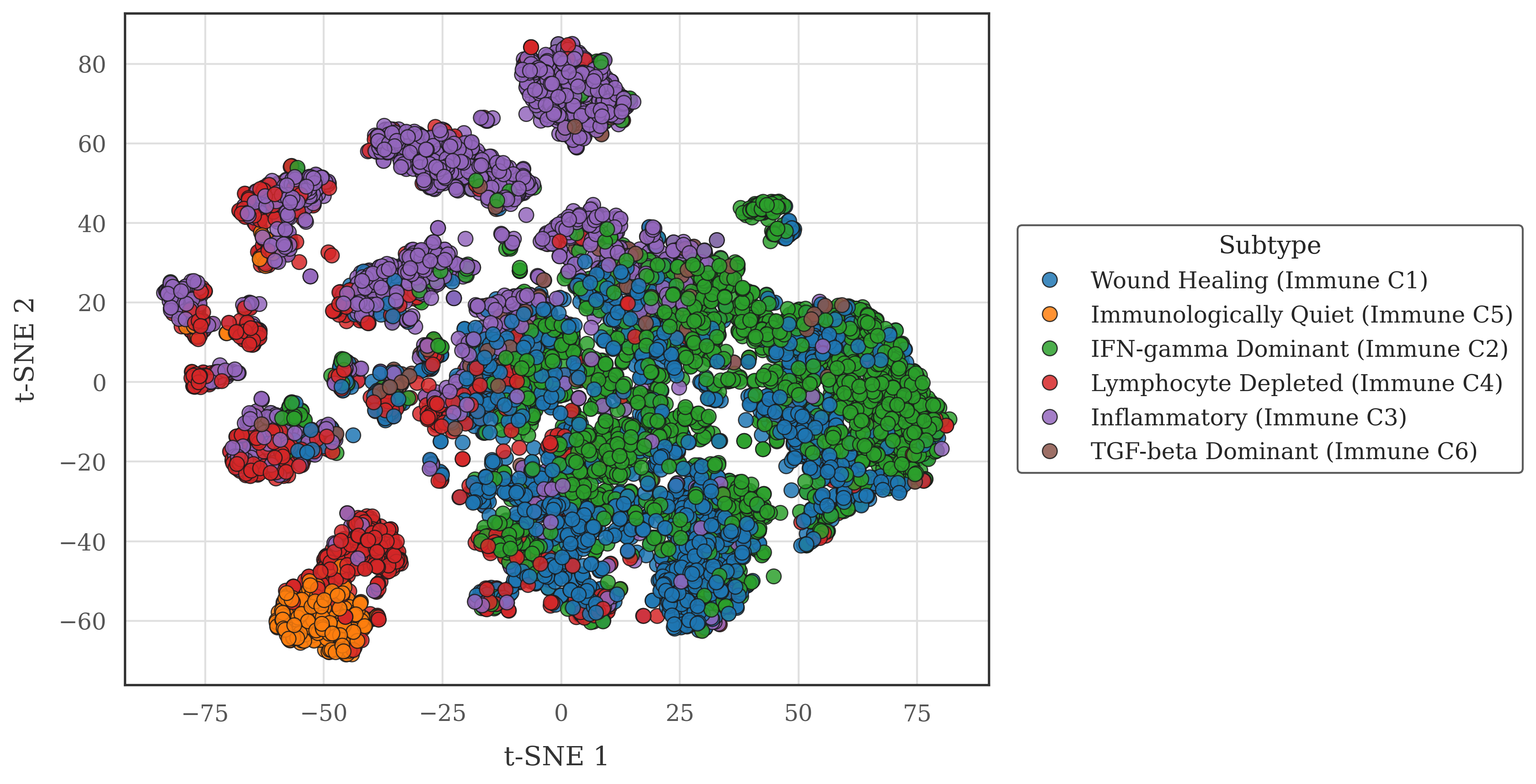}%
    }{%
      \fbox{\begin{minipage}[c][0.28\textheight][c]{\textwidth}
        \centering\small
        Placeholder:\\[4pt]\texttt{tsne\_xgboost\_genes.png}
      \end{minipage}}%
    }
    \caption{t-SNE visualization using XGBoost-selected features.}
    \label{fig:tsne_xgboost}
  \end{subfigure}
  \caption{Gene selection and clustering analysis. (a) t-SNE visualization of
    ANOVA-selected genes, and (b) t-SNE visualization using XGBoost-selected
    features.}
  \label{fig:clustering}
\end{figure}

The GMM algorithm was manually set to cluster the genes into 150 metagenes,
labeled Metagene 0 through Metagene 149. The mean number of genes per
cluster is 29.93, and the median is 25. However, the number of genes per cluster
varied greatly, with a standard deviation of 22.70 and a range of 113 (1 through
114). The distribution of gene counts across metagenes is shown in Figure 1a.
RFECV for the XGBoost model selected 68 features from 48 unique meta-
genes (out of 450 initial features), predominantly using mean values (49/68), in-
dicating that average gene expression within metagenes provided the strongest
signal for immune subtype classification. These features were used for further
interpretation. Figure 1b shows the t-SNE visualization of the ANOVA-selected
genes, while Figure 1c shows the t-SNE visualization of the XGBoost-selected
features.

\subsection{Predictive Model Performance}

\begin{table}[H]
\centering
\caption{Comparison of training and test set metrics across the three models.}
\label{tab:metrics}
\resizebox{\textwidth}{!}{%
\begin{tabular}{lcccccc}
\toprule
Metric & XGB Train & XGB Test & SVM Train & SVM Test & RF Train & RF Test \\
\midrule
Accuracy & 100.0\% $\pm$ 0.0\% & 87.8\% $\pm$ 0.9\% & 97.8\% $\pm$ 0.1\% & 86.5\% $\pm$ 0.5\% & 92.1\% $\pm$ 0.2\% & 82.5\% $\pm$ 1.1\% \\
Precision (Macro) & 100.0\% $\pm$ 0.0\% & 82.9\% $\pm$ 2.1\% & 98.1\% $\pm$ 0.2\% & 78.6\% $\pm$ 1.4\% & 87.3\% $\pm$ 0.5\% & 75.7\% $\pm$ 1.7\% \\
Precision (Weighted) & 100.0\% $\pm$ 0.0\% & 87.4\% $\pm$ 0.9\% & 97.8\% $\pm$ 0.1\% & 86.3\% $\pm$ 0.5\% & 93.0\% $\pm$ 0.1\% & 83.4\% $\pm$ 1.2\% \\
Recall (Macro) & 100.0\% $\pm$ 0.0\% & 78.1\% $\pm$ 1.2\% & 97.1\% $\pm$ 0.2\% & 77.8\% $\pm$ 1.1\% & 94.1\% $\pm$ 0.2\% & 77.3\% $\pm$ 2.2\% \\
Recall (Weighted) & 100.0\% $\pm$ 0.0\% & 87.8\% $\pm$ 0.9\% & 97.8\% $\pm$ 0.1\% & 86.5\% $\pm$ 0.5\% & 92.1\% $\pm$ 0.2\% & 82.5\% $\pm$ 1.1\% \\
F1 Score (Macro) & 100.0\% $\pm$ 0.0\% & 79.7\% $\pm$ 1.5\% & 97.6\% $\pm$ 0.2\% & 78.1\% $\pm$ 1.2\% & 89.4\% $\pm$ 0.5\% & 75.9\% $\pm$ 1.8\% \\
F1 Score (Weighted) & 100.0\% $\pm$ 0.0\% & 87.4\% $\pm$ 0.8\% & 97.8\% $\pm$ 0.1\% & 86.4\% $\pm$ 0.5\% & 92.3\% $\pm$ 0.2\% & 82.8\% $\pm$ 1.2\% \\
ROC AUC (Macro) & 100.0\% $\pm$ 0.0\% & 98.2\% $\pm$ 0.2\% & 99.9\% $\pm$ 0.0\% & 97.7\% $\pm$ 0.5\% & 99.5\% $\pm$ 0.0\% & 96.9\% $\pm$ 0.3\% \\
ROC AUC (Weighted) & 100.0\% $\pm$ 0.0\% & 98.5\% $\pm$ 0.2\% & 99.9\% $\pm$ 0.0\% & 98.3\% $\pm$ 0.1\% & 99.3\% $\pm$ 0.0\% & 97.1\% $\pm$ 0.2\% \\
\bottomrule
\end{tabular}%
}
\end{table}

Though all three models showed strong predictive performance, XGBoost demon-
strated the highest test accuracy (87.8\% $\pm$ 0.9\%), outperforming SVM (86.5\%
$\pm$ 0.5\%) and Random Forest (82.5\% $\pm$ 1.1\%), as shown in Table \ref{tab:metrics}.

\begin{figure}[ht]
  \centering
  \IfFileExists{confusion_matrix_XG.png}{%
    \includegraphics[width=0.9\textwidth]{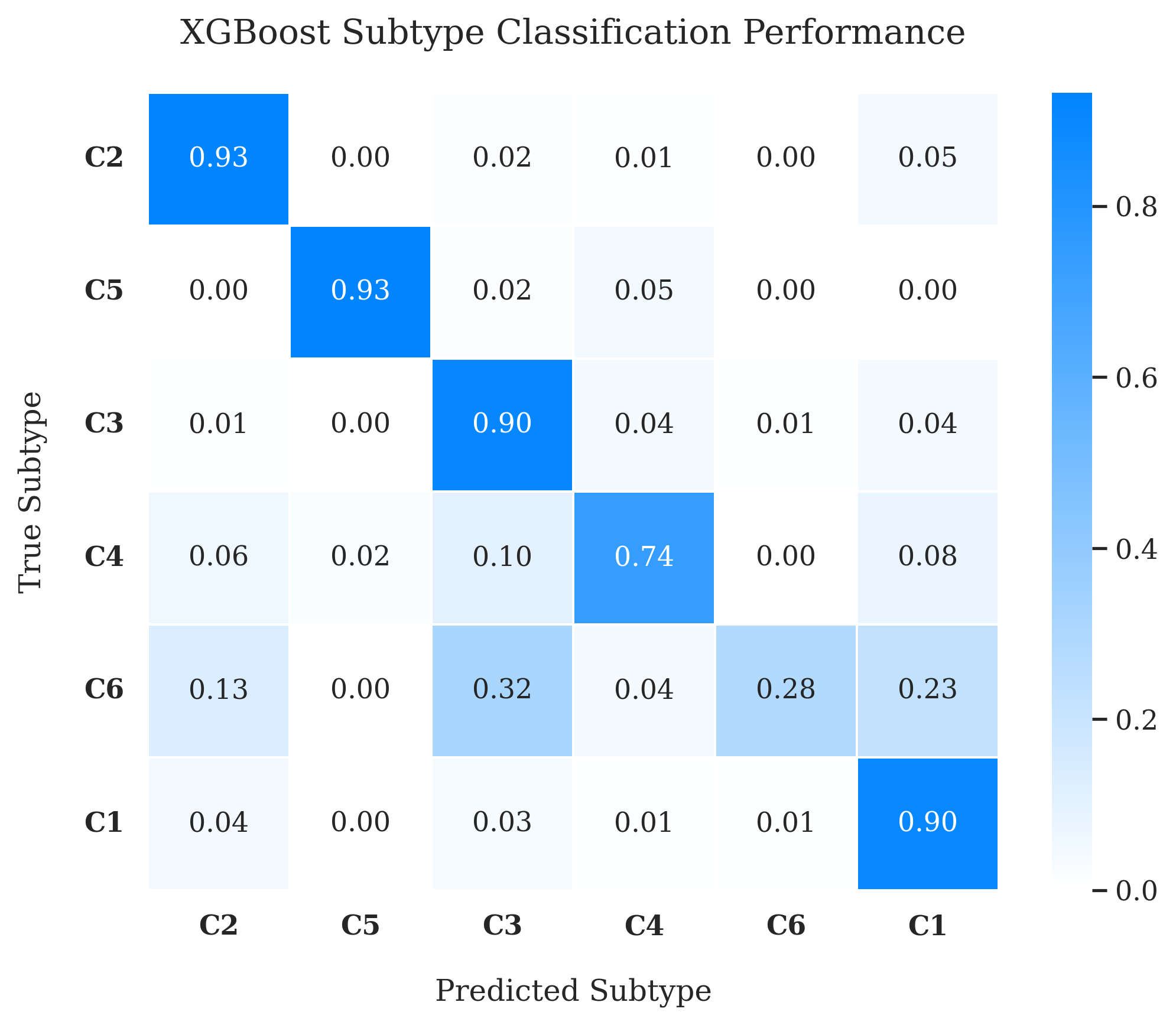}%
  }{%
    \fbox{\begin{minipage}[c][0.30\textheight][c]{0.85\textwidth}
      \centering\small
      Placeholder for figure file: \texttt{confusion\_matrix\_XG.png}\\[6pt]
      (Replace with the actual image file)
    \end{minipage}}%
  }
  \caption{Normalized confusion matrix of predicted versus true subtype labels for the XGBoost model.}
  \label{fig:confusion}
\end{figure}

XGBoost achieved strong performance for subtypes C1, C2, C3, and C5 (\(\geq\)
90\% recall), moderate performance for C4 (74\% recall), but poor performance
for C6 (28\% recall). C6 was frequently misclassified as C3 (32\%), C1 (23\%), or
C2 (12\%) (Figure \ref{fig:confusion}). All models showed lower performance with macro metrics
compared to weighted metrics.

\subsection{Feature Importance and Interpretability Analysis}

\begin{table}[H]
\centering
\caption{Top ten most predictive metagenes from SHAP analysis. The number of subtypes in which a metagene appeared and the total number of appearances across summary statistics were calculated, considering only the top five metagene-derived features per subtype.}
\label{tab:top_metagenes}
\begin{tabular}{lcc}
\toprule
Rank & Metagene & \# Subtypes / \# Appearances \\
\midrule
1 & Metagene 82 & 4 / 8 \\
2 & Metagene 147 & 3 / 3 \\
3 & Metagene 87 & 3 / 3 \\
4 & Metagene 27 & 2 / 3 \\
5 & Metagene 59 & 2 / 2 \\
6 & Metagene 55 & 2 / 2 \\
7 & Metagene 13 & 2 / 2 \\
8 & Metagene 45 & 1 / 1 \\
9 & Metagene 68 & 1 / 1 \\
10 & Metagene 81 & 1 / 1 \\
\bottomrule
\end{tabular}
\end{table}

\begin{figure}[ht]
  \centering
  \begin{subfigure}[t]{0.31\textwidth}
    \centering
    \IfFileExists{SHAP_Figures/C1_SHAP.png}{%
      \includegraphics[width=\textwidth]{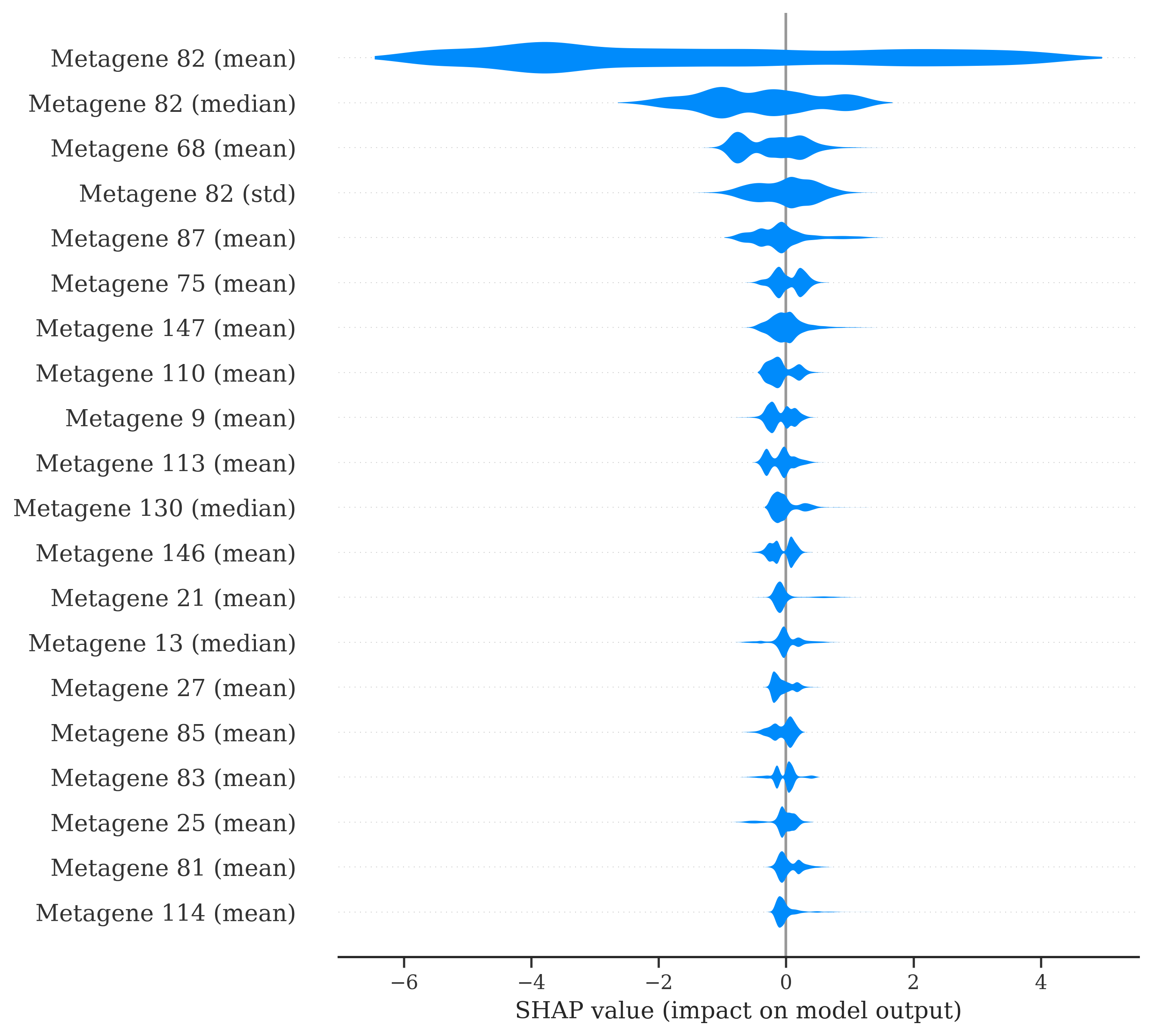}%
    }{%
      \fbox{\begin{minipage}[c][0.22\textheight][c]{\textwidth}
        \centering\small Placeholder:\\\texttt{C1\_SHAP.png}
      \end{minipage}}%
    }
    \caption{Subtype C1}
    \label{fig:shap_c1}
  \end{subfigure}
  \hfill
  \begin{subfigure}[t]{0.31\textwidth}
    \centering
    \IfFileExists{SHAP_Figures/C2_SHAP.png}{%
      \includegraphics[width=\textwidth]{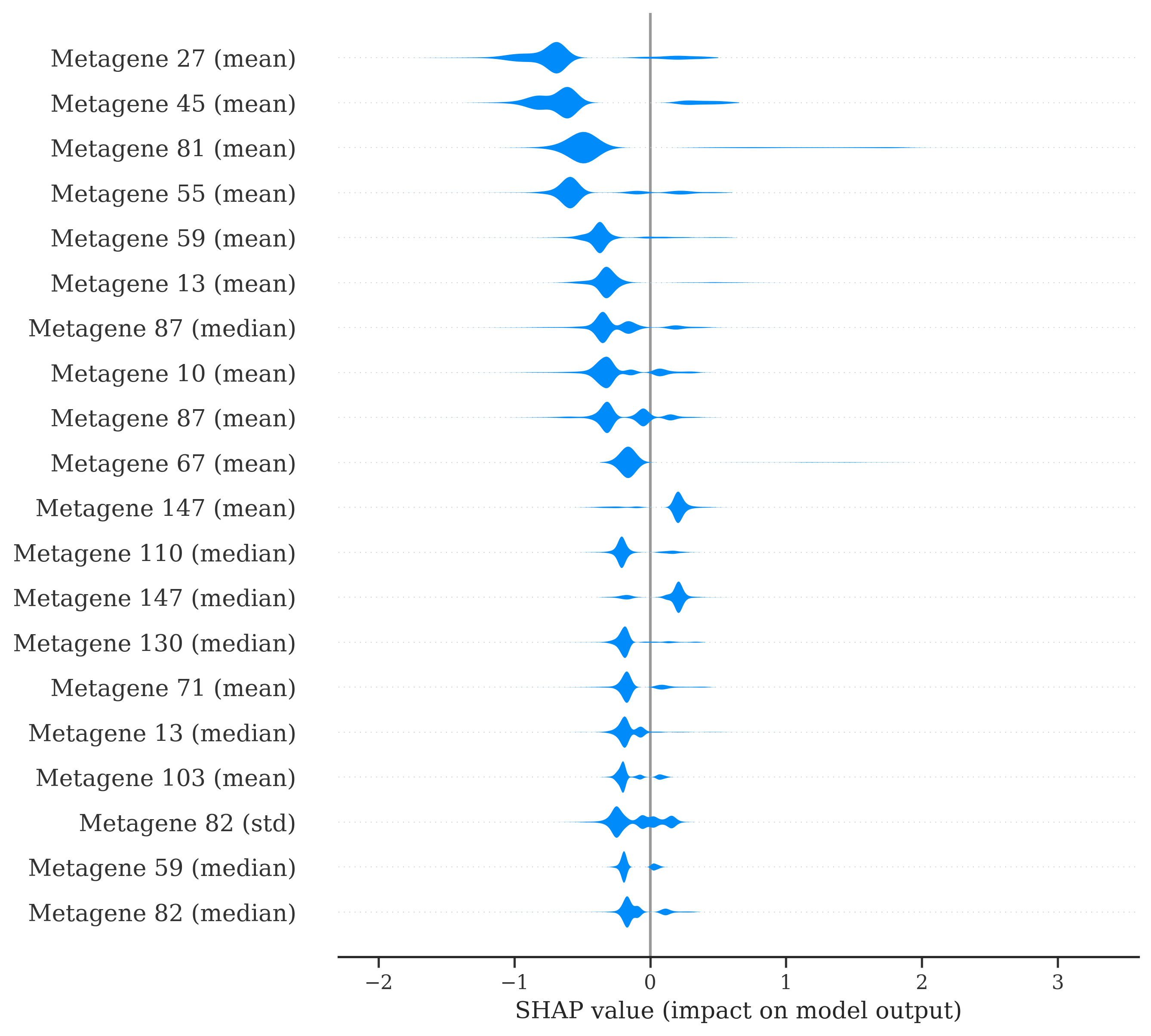}%
    }{%
      \fbox{\begin{minipage}[c][0.22\textheight][c]{\textwidth}
        \centering\small Placeholder:\\\texttt{C2\_SHAP.png}
      \end{minipage}}%
    }
    \caption{Subtype C2}
    \label{fig:shap_c2}
  \end{subfigure}
  \hfill
  \begin{subfigure}[t]{0.31\textwidth}
    \centering
    \IfFileExists{SHAP_Figures/C3_SHAP.png}{%
      \includegraphics[width=\textwidth]{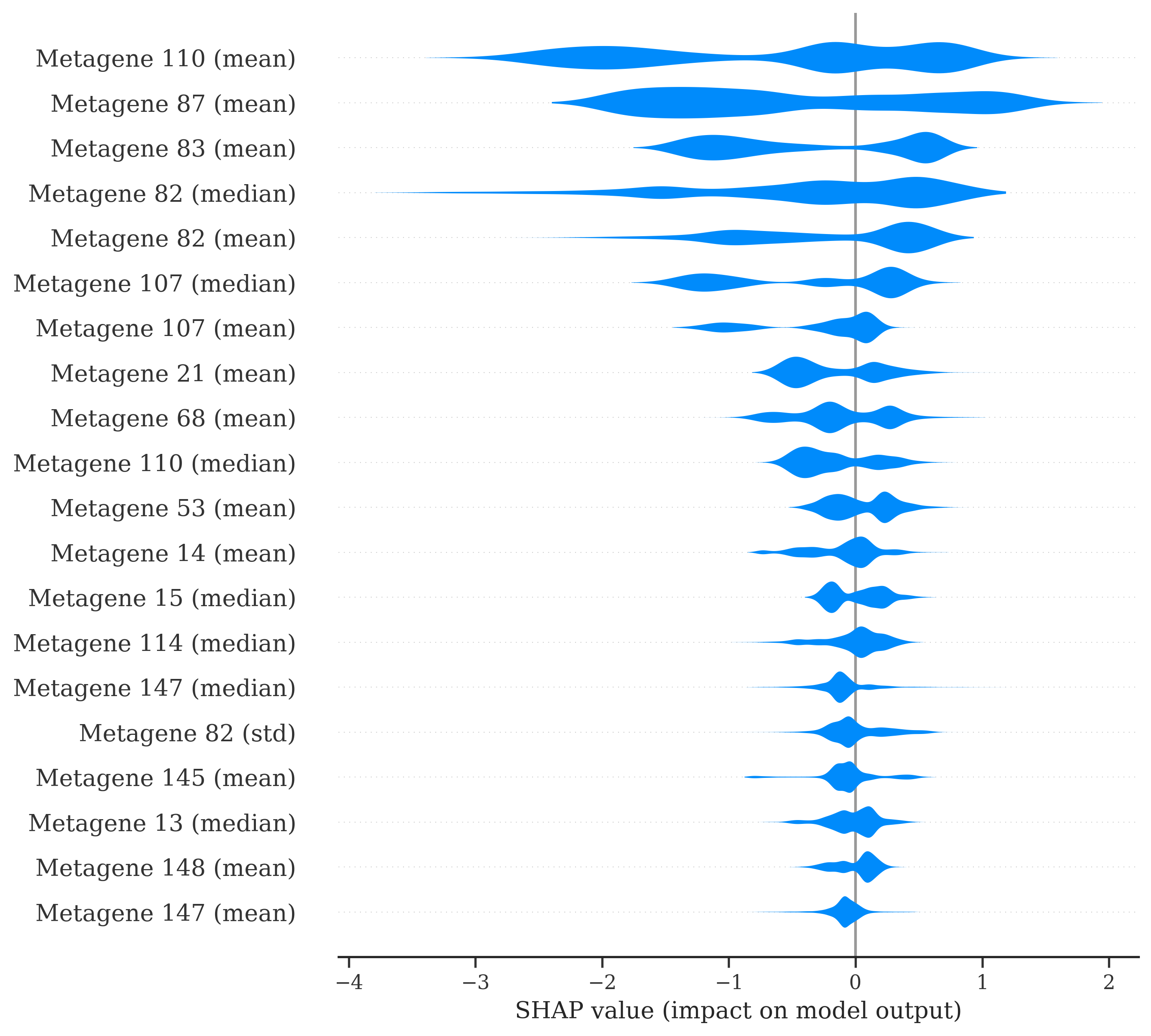}%
    }{%
      \fbox{\begin{minipage}[c][0.22\textheight][c]{\textwidth}
        \centering\small Placeholder:\\\texttt{C3\_SHAP.png}
      \end{minipage}}%
    }
    \caption{Subtype C3}
    \label{fig:shap_c3}
  \end{subfigure}

  \vspace{0.75em}

  \begin{subfigure}[t]{0.31\textwidth}
    \centering
    \IfFileExists{SHAP_Figures/C4_SHAP.png}{%
      \includegraphics[width=\textwidth]{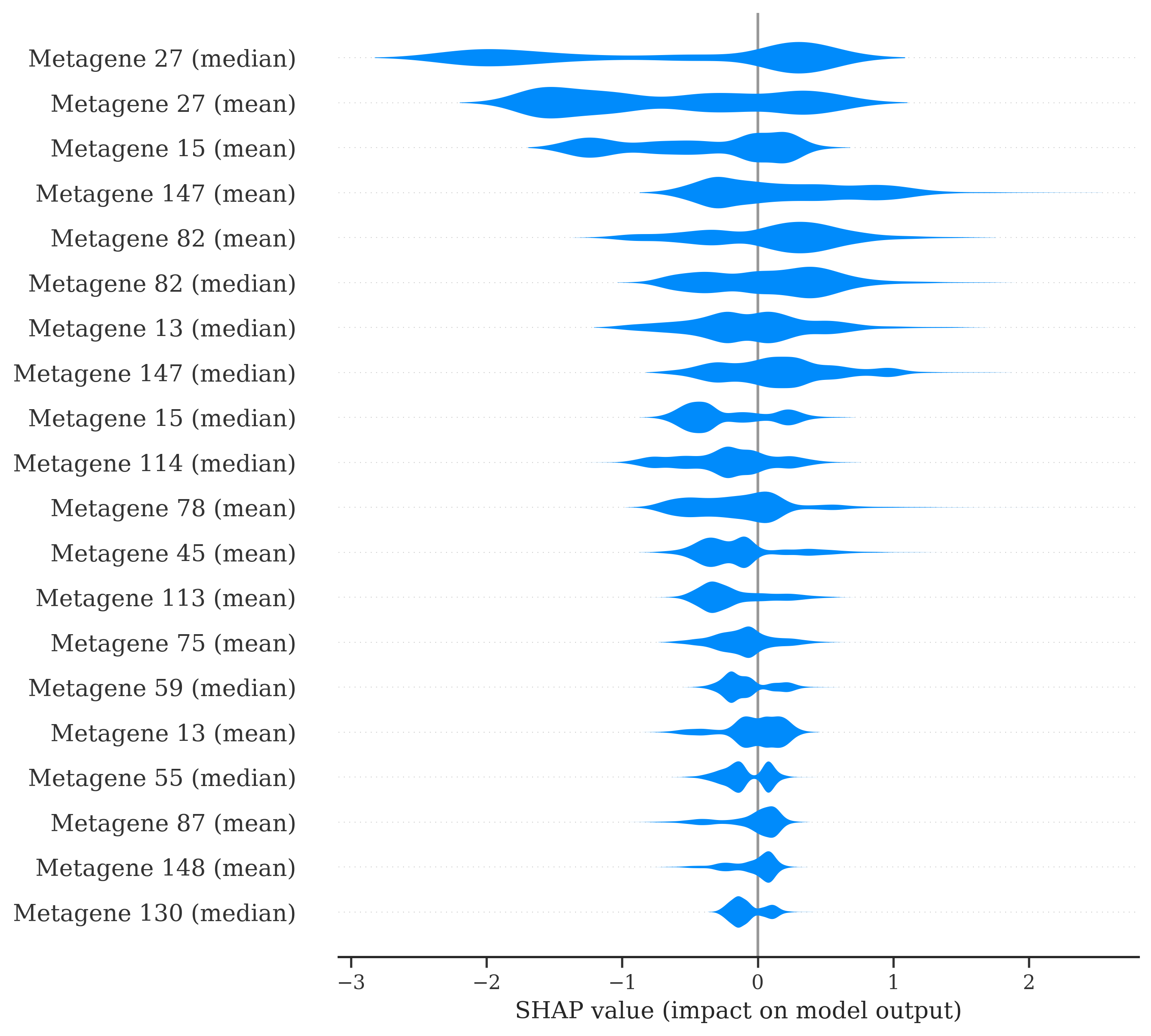}%
    }{%
      \fbox{\begin{minipage}[c][0.22\textheight][c]{\textwidth}
        \centering\small Placeholder:\\\texttt{C4\_SHAP.png}
      \end{minipage}}%
    }
    \caption{Subtype C4}
    \label{fig:shap_c4}
  \end{subfigure}
  \hfill
  \begin{subfigure}[t]{0.31\textwidth}
    \centering
    \IfFileExists{SHAP_Figures/C5_SHAP.png}{%
      \includegraphics[width=\textwidth]{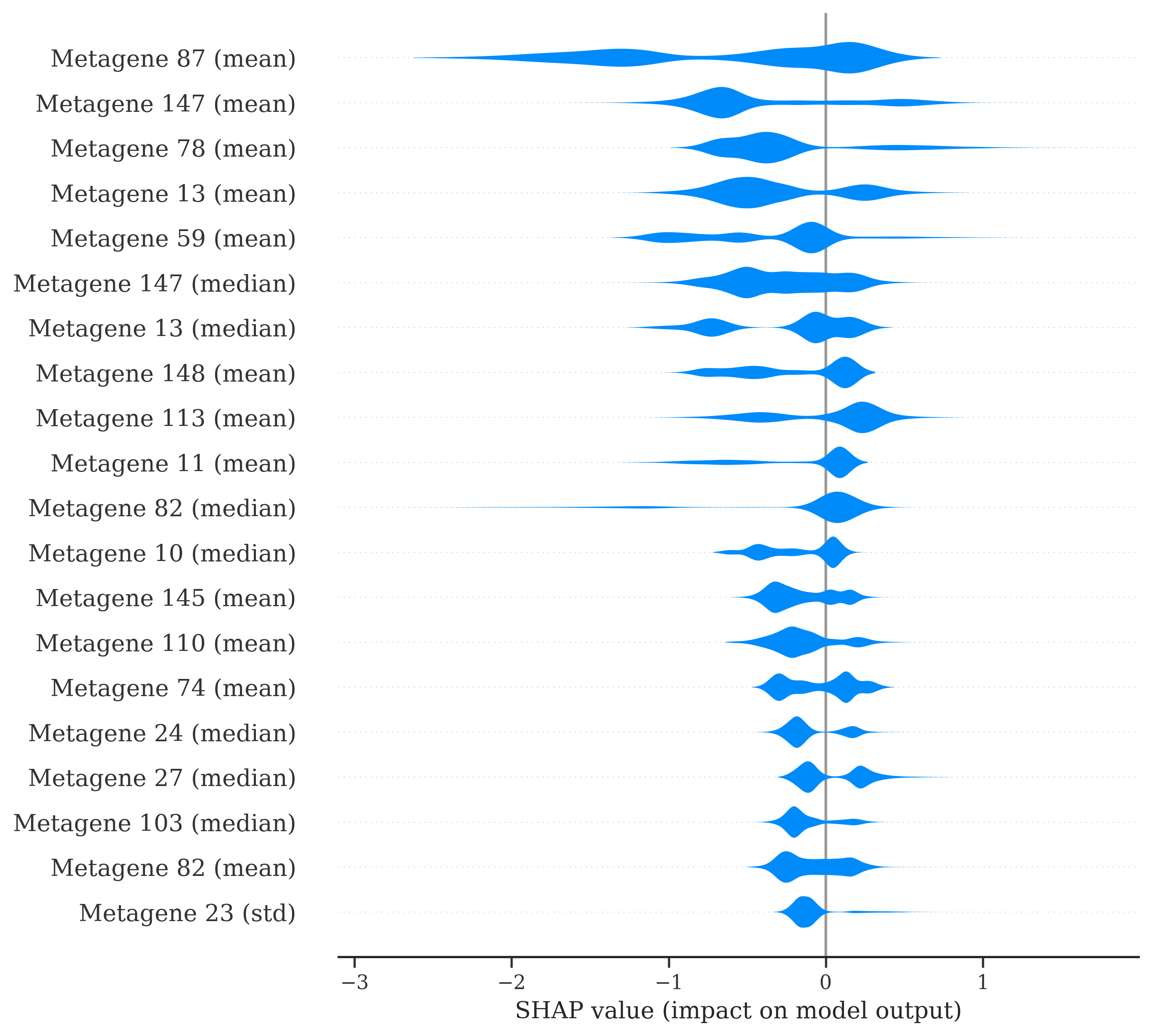}%
    }{%
      \fbox{\begin{minipage}[c][0.22\textheight][c]{\textwidth}
        \centering\small Placeholder:\\\texttt{C5\_SHAP.png}
      \end{minipage}}%
    }
    \caption{Subtype C5}
    \label{fig:shap_c5}
  \end{subfigure}
  \hfill
  \begin{subfigure}[t]{0.31\textwidth}
    \centering
    \IfFileExists{SHAP_Figures/C6_SHAP.png}{%
      \includegraphics[width=\textwidth]{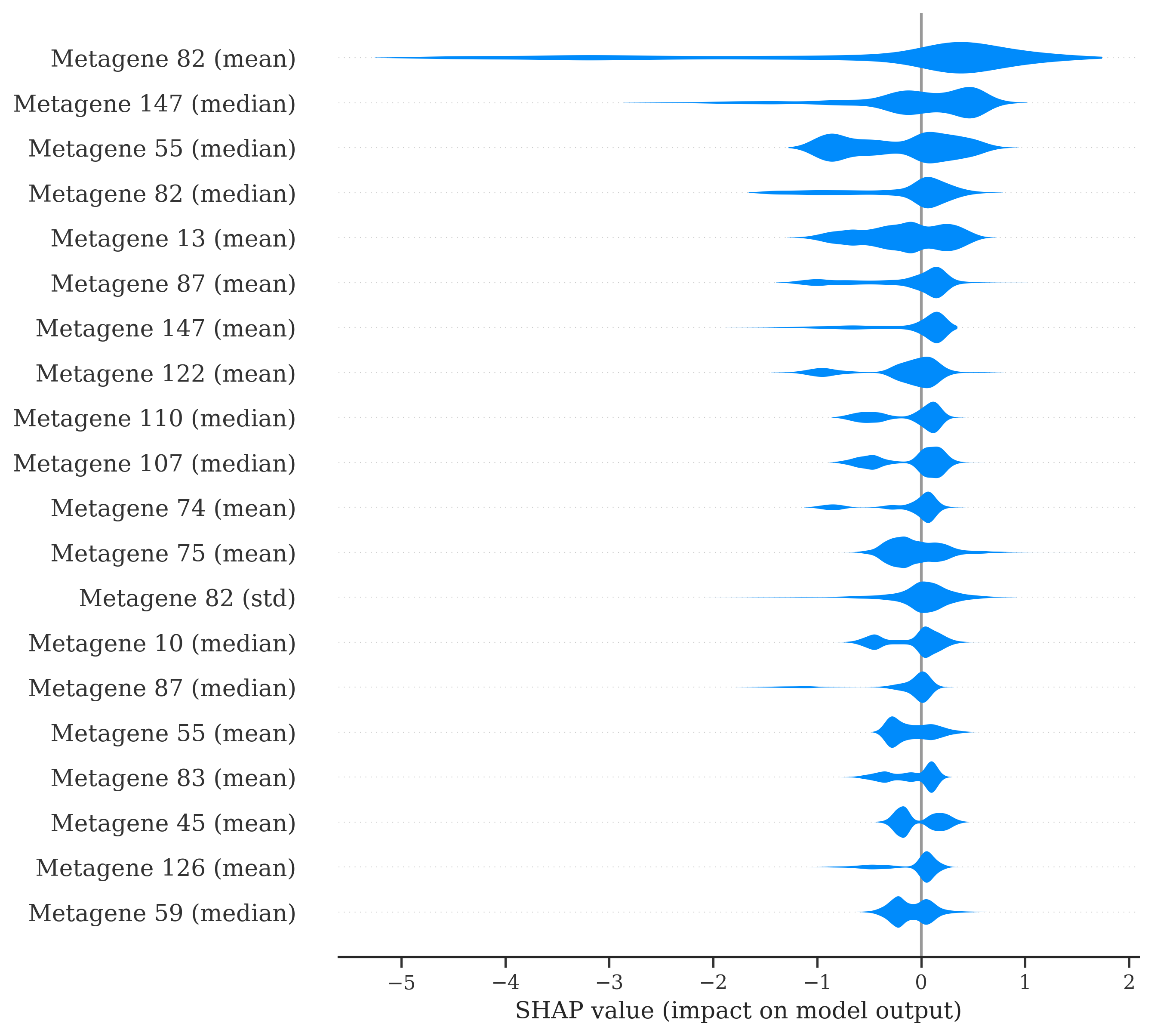}%
    }{%
      \fbox{\begin{minipage}[c][0.22\textheight][c]{\textwidth}
        \centering\small Placeholder:\\\texttt{C6\_SHAP.png}
      \end{minipage}}%
    }
    \caption{Subtype C6}
    \label{fig:shap_c6}
  \end{subfigure}

  \caption{Panels a--f (corresponding to subtypes C1--C6) show SHAP value
    distributions for the top 20 most predictive metagene-derived features
    identified for each subtype. Each violin represents the contribution of a
    feature to the model's prediction, with wider sections indicating higher
    sample density. Positive and negative SHAP values reflect feature influence
    toward or against classification of the corresponding subtype.}
  \label{fig:shap}
\end{figure}

SHAP analysis revealed three metagenes as the most predictive overall: Metagene 82 (top 5 across four subtypes), Metagene 87 (top 5 across three subtypes), and Metagene 147 (top 5 across three subtypes). The top ten most predictive metagenes are listed in Table \ref{tab:top_metagenes}, while the SHAP violin plots in Figure \ref{fig:shap} summarize the top 20 metagene-derived features for each subtype.

\subsection{Survival Analysis and Prognostic Significance}

\subsubsection{Kaplan-Meier Analysis}

\begin{figure}[ht]
  \centering
  \begin{subfigure}[t]{0.31\textwidth}
    \centering
    \IfFileExists{KM_Metagene_110_mean.png}{%
      \includegraphics[width=\textwidth]{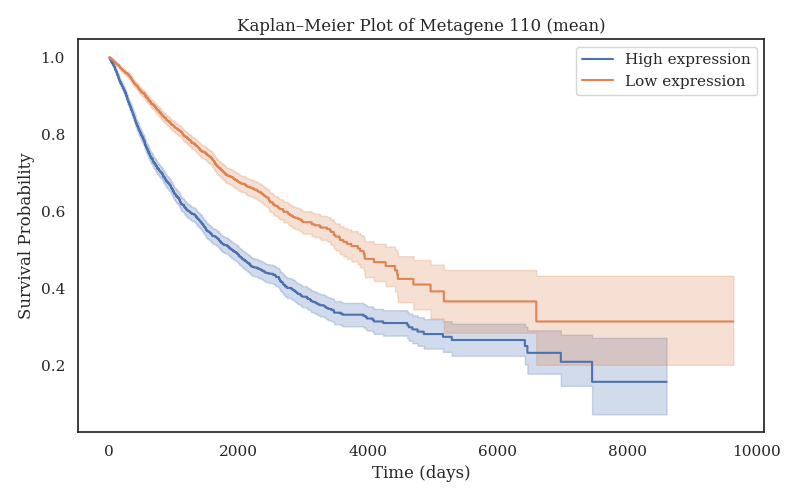}%
    }{%
      \fbox{\begin{minipage}[c][0.22\textheight][c]{\textwidth}
        \centering\small Placeholder:\\\texttt{KM\_Metagene\_110\_mean.png}
      \end{minipage}}%
    }
    \caption{Metagene 110 (mean)}
    \label{fig:km_110}
  \end{subfigure}
  \hfill
  \begin{subfigure}[t]{0.31\textwidth}
    \centering
    \IfFileExists{KM_Metagene_81_mean.png}{%
      \includegraphics[width=\textwidth]{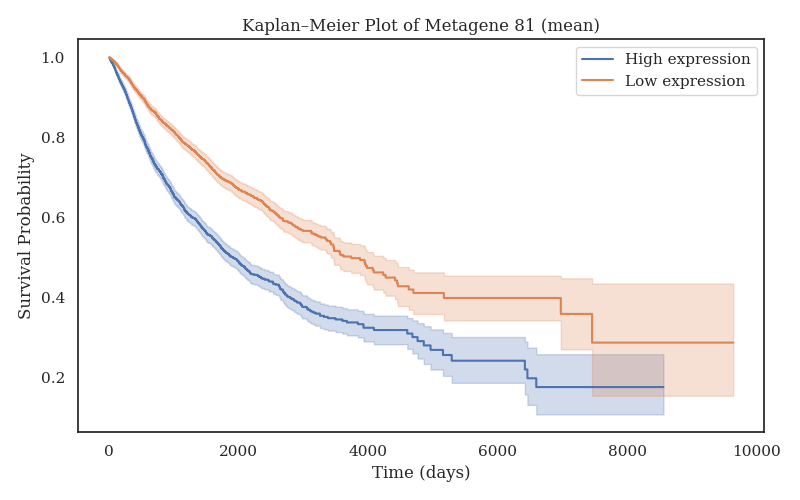}%
    }{%
      \fbox{\begin{minipage}[c][0.22\textheight][c]{\textwidth}
        \centering\small Placeholder:\\\texttt{KM\_Metagene\_81\_mean.png}
      \end{minipage}}%
    }
    \caption{Metagene 81 (mean)}
    \label{fig:km_81}
  \end{subfigure}
  \hfill
  \begin{subfigure}[t]{0.31\textwidth}
    \centering
    \IfFileExists{KM_Metagene_101_mean.png}{%
      \includegraphics[width=\textwidth]{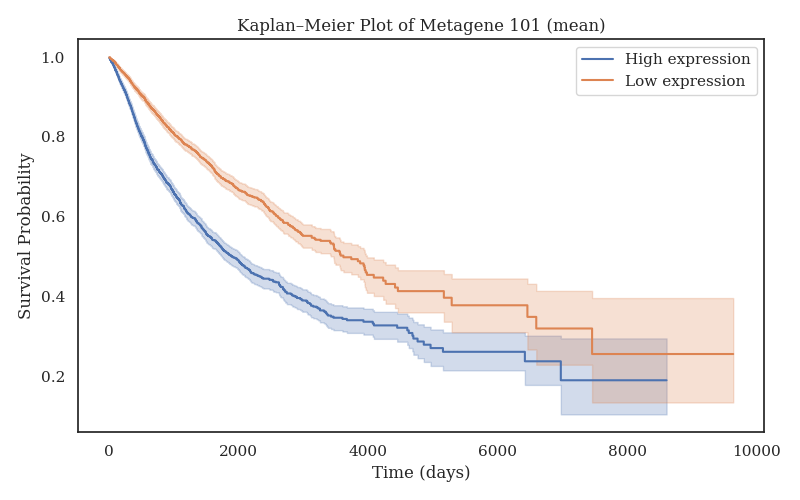}%
    }{%
      \fbox{\begin{minipage}[c][0.22\textheight][c]{\textwidth}
        \centering\small Placeholder:\\\texttt{KM\_Metagene\_101\_mean.png}
      \end{minipage}}%
    }
    \caption{Metagene 101 (mean)}
    \label{fig:km_101}
  \end{subfigure}
  \caption{Kaplan--Meier plots showing overall survival differences between
    high (above-median) and low (below-median) expression groups for three
    distinct metagene-derived features: (a) Metagene 110 (mean), (b) Metagene
    81 (mean), and (c) Metagene 101 (mean).}
  \label{fig:km}
\end{figure}

Kaplan-Meier analysis of the 68 XGBoost features revealed that 56 features (40 metagenes) were significantly associated (FDR < 0.05) with survival outcomes after BH correction. Of the top 20 features, most (85\%) showed worse survival when overexpressed. Twenty-nine features demonstrated clinically meaningful survival differences between high- and low-expression groups (\(\geq\)10\%), with Metagenes 110, 81, and 101 showing the largest differences (\(\sim\)18\%) (see Table \ref{tab:km_top}). Figure \ref{fig:km} shows Kaplan-Meier survival curves for the mean features of Metagene 110, 81, and 101. All Kaplan--Meier results, including plots, are presented in the supplementary materials.

\begin{table}[H]
\centering
\caption{Top metagene-derived features ranked by survival significance. The table reports 5-year survival percentages for each group, the absolute survival difference, and the adjusted p-value from Kaplan--Meier analysis (ranked by lowest adjusted p-value).}
\label{tab:km_top}
\begin{tabular}{lccccc}
\toprule
Metagene & Stat & Low (\%) & High (\%) & $\Delta$ 5-yr (\%) & Adj. p \\
\midrule
Metagene 110 & mean & 50.9 & 69.3 & 18.4 & 2.38e-64 \\
Metagene 110 & median & 51.0 & 69.2 & 18.2 & 2.37e-57 \\
Metagene 81 & mean & 51.0 & 68.9 & 17.9 & 2.68e-53 \\
Metagene 101 & mean & 50.9 & 68.8 & 17.9 & 1.98e-52 \\
Metagene 14 & mean & 51.8 & 68.5 & 16.7 & 6.33e-50 \\
Metagene 107 & median & 52.2 & 68.0 & 15.8 & 2.29e-47 \\
Metagene 146 & mean & 51.9 & 68.2 & 16.3 & 4.59e-47 \\
Metagene 107 & mean & 52.3 & 67.8 & 15.5 & 1.61e-45 \\
Metagene 85 & mean & 52.3 & 67.9 & 15.6 & 1.45e-44 \\
Metagene 146 & median & 52.1 & 68.0 & 15.9 & 8.11e-42 \\
Metagene 108 & mean & 51.9 & 67.9 & 16.0 & 6.79e-40 \\
Metagene 75 & mean & 52.8 & 67.6 & 14.8 & 3.39e-38 \\
\bottomrule
\end{tabular}
\end{table}

\subsubsection{Cox Regression with Clinical Covariates}

Multivariate Cox Regression with the 68 XGBoost features, cancer type, and age at diagnosis as covariates identified six metagenes with significant survival associations after adjusting for clinical confounders. Four metagenes (75, 14, 107, 78) predicted worse survival (HR>1), while two (133, 73) predicted improved survival (HR<1), as shown in Table \ref{tab:cox}.

Among clinical variables, age at diagnosis showed the most statistically significant association with survival, although it had a small effect size (HR = 1.017). Cancer type showed high but varying effect sizes. Complete Cox regression results are provided in the supplementary materials.

\begin{table}[H]
\centering
\caption{Hazard ratios for top metagene-derived features. The table reports hazard ratios (HR) with 95\% confidence intervals and adjusted p-values from Cox regression.}
\label{tab:cox}
\begin{tabular}{lcccc}
\toprule
Metagene & Stat & HR & 95\% CI & Adj. p \\
\midrule
Metagene 75 & mean & 1.099 & 1.042 - 1.160 & 3.84e-03 \\
Metagene 14 & mean & 1.051 & 1.021 - 1.083 & 5.89e-03 \\
Metagene 107 & median & 1.055 & 1.022 - 1.089 & 5.89e-03 \\
Metagene 107 & mean & 1.055 & 1.021 - 1.090 & 7.29e-03 \\
Metagene 78 & mean & 1.081 & 1.023 - 1.142 & 2.76e-02 \\
Metagene 133 & median & 0.949 & 0.915 - 0.986 & 2.76e-02 \\
Metagene 73 & mean & 0.983 & 0.970 - 0.996 & 4.59e-02 \\
\bottomrule
\end{tabular}
\end{table}

\subsection{Functional Characterization of the Predictive Metagenes}

Functional enrichment analysis using g:Profiler revealed that 12 of the 38 predictive metagenes identified by SHAP were significantly enriched (FDR < $10^{-9}$) for biological processes (Table \ref{tab:enrichment}). These processes fell into three general categories: cell cycle regulation (4 metagenes), immune activation (4), and ECM organization (3), with Metagenes 21, 82, and 114 showing the strongest associations.

Of the 48 metagenes, the vast majority (44) had all their genes upregulated. The genes in Metagenes 61 and 110 were overwhelmingly upregulated (approximately 98\% and 91\% of genes, respectively). In contrast, Metagenes 23 and 84 were fully and mostly downregulated, respectively. However, these two metagenes only had a standard deviation summary statistic, which likely caused the high metagene score group to not reflect a higher metagene expression value. Table \ref{tab:rep_metagenes} shows the top 3 upregulated genes and their expression values for six representative metagenes.

\begin{table}[H]
\centering
\caption{Top metagenes ranked by minimum adjusted p-value and their associated biological pathways.}
\label{tab:enrichment}
\begin{tabular}{lll}
\toprule
Rank & Metagene & Biological Pathways (Min adj p-value) \\
\midrule
1 & Metagene 21 & Cell cycle process (6.4e-37) \\
2 & Metagene 82 & Defense response to virus (1.1e-35) \\
3 & Metagene 114 & T-cell activation / MHC I binding (9.2e-32) \\
4 & Metagene 107 & Cell cycle process (3.4e-30) \\
5 & Metagene 85 & Cell cycle process (4.0e-26) \\
6 & Metagene 15 & T-cell activation (3.9e-24) \\
7 & Metagene 147 & Immune response / MHC II binding (3.5e-23) \\
8 & Metagene 59 & Collagen-containing ECM (5.6e-17) \\
9 & Metagene 55 & ECM Organization (1.3e-14) \\
10 & Metagene 78 & ECM Organization (1.3e-12) \\
11 & Metagene 83 & DNA Metabolic Process / Cell cycle (3.6e-12) \\
12 & Metagene 25 & Tube development / Angiogenesis (6.4e-10) \\
\bottomrule
\end{tabular}
\end{table}

\begin{table}[H]
\centering
\caption{Representative metagenes with top upregulated genes and mean expression values.}
\label{tab:rep_metagenes}
\begin{tabular}{lllr}
\toprule
Metagene & Num.\ Genes & Top Genes & Expr.\ (Mean) \\
\midrule
\multirow{3}{*}{Metagene 147} & \multirow{3}{*}{51} & HLA-DQA1 & 4.52 \\
 &  & IL2RA & 4.43 \\
 &  & HLA-DQA2 & 4.38 \\
\midrule
\multirow{3}{*}{Metagene 82} & \multirow{3}{*}{41} & OASL & 4.27 \\
 &  & ZBP1 & 4.16 \\
 &  & IFI44L & 4.03 \\
\midrule
\multirow{3}{*}{Metagene 85} & \multirow{3}{*}{43} & MYBL2 & 6.19 \\
 &  & UBE2C & 5.52 \\
 &  & AURKB & 5.41 \\
\midrule
\multirow{3}{*}{Metagene 107} & \multirow{3}{*}{49} & DLGAP5 & 5.65 \\
 &  & NEK2 & 5.31 \\
 &  & CENPA & 5.22 \\
\midrule
\multirow{3}{*}{Metagene 75} & \multirow{3}{*}{32} & PLAU & 4.29 \\
 &  & PLAUR & 3.92 \\
 &  & LIF & 3.65 \\
\midrule
\multirow{3}{*}{Metagene 78} & \multirow{3}{*}{17} & COL6A2 & 4.02 \\
 &  & BGN & 3.49 \\
 &  & COL4A1 & 2.89 \\
\bottomrule
\end{tabular}
\end{table}

\section{Discussion}

\subsection{Interpretation of Pan-Cancer Immune Metagenes}

Through clustering, SHAP, enrichment, and survival analysis, 48 distinct metagenes were identified across a diverse pan-cancer cohort, providing a novel, data-driven framework for characterizing the complexity of the tumor immune microenvironment. These metagenes, unlike single-gene biomarkers, represent coordinated immune and tumor intrinsic programs, offering a more robust and biologically meaningful representation of immune states.

The metagenes identified transcriptional programs that not only predicted immune subtype, but also reflected well-established biological processes in cancer. Representative metagenes were identified from three major categories, and their gene contents were examined in the context of existing literature.

\subsubsection{Coordinated Immune Activation Programs}

Our analysis revealed that immune activation in tumors involves coordinated expression of multiple complementary pathways rather than individual immune-related genes. Metagenes 82 and 147, predictive across multiple immune subtypes, simultaneously demonstrated interferon signaling (Metagene 82), lymphocyte activation, and antigen-presentation (Metagene 147). Metagene 147 was dominated by HLA genes, representing strong antigen presentation via MHC II, key for effective anti-tumor responses. However, this was balanced by genes such as FCGR2B, PD-L2, and LILRBs with documented roles in tumor immune suppression \cite{gong2022,chuang2024,wang2022,deng2021}.

Metagene 82, on the other hand, was dominated by type I interferon-stimulated genes like OASL, IFI27, and RSAD2, which have been shown to have dual roles in the tumor immune microenvironment. Although interferons are critical regulators of anti-tumor immune responses, chronic IFN signaling can promote immune evasion and immunotherapy resistance in tumors. OASL, for example, activates signaling pathways including mTORC1, IFN-\(\gamma\)/STAT1, and JAK/STAT3, upregulating immune checkpoint ligands and enhancing immunosuppressive signaling. Similarly, IFI27 is associated with decreased CD8+ T-cell infiltration and increased M2-like macrophage infiltration \cite{boukhaled2020,liu2024,zhao2023,shojaei2025}.

These two transcriptional modules reveal that tumor immune activation involves context-dependent mechanisms and a balance between pro- and anti-tumor immunity. Traditional immune biomarkers often focus on individual markers like PD-1 expression or CD8+ T-cell counts. However, our analysis reveals that activation markers are often co-expressed with suppressive elements within the same transcriptional program. This coordination suggests that effective immunotherapy may require combination strategies that enhance immune activation while simultaneously blocking the co-expressed suppressive mechanisms within the same program.

\subsubsection{Tumor Intrinsic Programs with Immune Consequences}

We identified tumor intrinsic programs of cell cycle progression and regulation that link to immune evasion mechanisms. Metagenes 85 and 107 were strongly enriched for similar biological programs. Regulators of the G2/M (MYBL2, CDK1) and M checkpoints (AURKB, UBE2C, TTK) were abundant across both metagenes; when upregulated, these genes drive cancer cell proliferation. Interestingly, some of these genes also influence the tumor immune microenvironment. Aurora Kinase B, for example, is associated with leukocyte infiltration, as well as increased CD4+ Th2 differentiation, contributing to both pro- and anti-tumor immune responses. Overexpression of UBE2C leads to not only cell cycle dysregulation, but is also positively associated with immune checkpoint-related genes such as PD-1 and CTLA-4, suggesting a role in immune evasion \cite{musa2017,ren2022,xie2016,li2024,jalali2024}.

Co-expressed with these are genes involved in DNA replication and repair.
Like checkpoint regulatory genes, upregulation of these genes is implicated in tumor progression and chemotherapy resistance \cite{alleramoreau2012,broustas2014}. These genes may also have roles in immune evasion, both directly and indirectly. Upregulation of RAD51, for example, leads to aberrant repair mechanisms while also displaying a strong correlation with immune checkpoint molecules and inflammatory cytokines \cite{liao2022}. Furthermore, genes such as CDC6, CDC20, and ORC6 represent a controlled tumor mutational burden, possibly contributing to the development of more immune-resistant subpopulations. CDC6, CDC20, and ORC6 are associated with both genomic instability and an immunosuppressive microenvironment \cite{pu2024,gayyed2015,wu2021,lin2023}.

Metagenes 85 and 107 show that tumor proliferation and immune evasion are interconnected processes. Many checkpoint regulators show strong associations with immune checkpoint expression and immunosuppressive microenvironments, while DNA replication and repair genes maintain a controlled level of genomic instability, which may indirectly support immune-resistant clones. This suggests that disrupting cell cycle progression could simultaneously restore immune sensitivity---supporting emerging therapeutic strategies employing CDK inhibitors with immunotherapies to overcome this coordinated resistance mechanism \cite{he2025}.

\subsubsection{Microenvironment Remodeling Networks}

In addition to immune activation and cell cycle regulation, our analysis also revealed coordinated ECM remodeling for immune exclusion, rather than isolated matrix changes. Metagene 78 represents a single transcriptional module of ECM strengthening and breakdown. It co-expresses collagen genes (COL4A1, COL6A2) that form a dense basement membrane structure with MMP14, a matrix metalloproteinase that clears the extracellular matrix to facilitate angiogenesis. These contradictory pathways reflect a dynamic ECM remodeling signature: enhanced stiffness limits T-cell infiltration and activates invasion pathways (e.g, TGF-\(\beta\)), while concurrent MMP14 expression permits controlled matrix degradation for vascular access and metastasis \cite{gatseva2019,kuczek2019,song2022,minde2021}. This physical matrix restructuring is complemented by matrix-associated factors like biglycan (BGN), which scaffolds collagen fibers while its soluble form activates inflammatory NF-\(\kappa\)B signaling in immune cells, and perlecan (HSPG2), which stabilizes basement membranes while interacting with growth factors to promote proliferation and invasion \cite{hayes2022,whitelock2008,appunni2021}.

Metagene 75 demonstrates how effective immune exclusion integrates multiple complementary pathways. This metagene coordinates both ECM crosslinking (PLAU) and degradation (LOXL2) alongside metabolic reprogramming through MTC4 (SLC16A3), which exports lactate from highly glycolytic cancer cells, creating an acidic microenvironment that suppresses cytotoxic T-cell and NK cell function \cite{cuevas2017,mekkawy2014,du2025,mortazavi2023}. Simultaneously, the program directly suppresses immune function through Leukemia Inhibitory Factor, which recruits TAMs and excludes CD8+ T-cells. Some genes have complex and context-dependent roles in tumor progression. Thrombospondin 1 (THBS1), by activating CD74, can inhibit angiogenesis, but may also suppress cytotoxic T-cell and NK cell function through CD47's role as a ``don't eat me'' signal on tumor cells \cite{kaur2021}. Additionally, SOCS3 can inhibit proliferation and invasion in many cancers, while simultaneously recruiting immunosuppressive cells \cite{dai2021}.

The strong survival associations of these metagenes in Cox regression suggest that these coordinated physical-metabolic-immune remodeling networks represent a fundamental mechanism of tumor progression beyond individual matrix alterations. This points towards multi-modal therapeutic strategies simultaneously targeting these microenvironment components.

\subsection{Clinical Significance and Predictive Power}

Thorsson et al. \cite{thorsson2019} have shown that the six immune subtypes have known relevance to the tumor immune landscape. The XGBoost model classifies a patient's immune subtype across a transcriptomic space spanning 48 unique metagenes, displaying high accuracy in predicting immune subtype with the metagene-derived features. This enables transcriptomic immune profiling without the need for external immune deconvolution tools.

More than just representing transcriptional modules of the tumor immune microenvironment, Cox regression and Kaplan-Meier analysis showed that many of the metagenes had statistically significant associations with survival outcomes. This suggests that more than just classifying immune subtype, the model and its predictive metagenes capture information relevant to patient outcomes.

Metagene-based biomarkers possess an advantage over single-gene biomarkers, as they reflect various biological processes and tumor heterogeneity, increasing their robustness and interpretability \cite{passaro2024}. The metagenes with both strong associations with survival outcomes and biological relevance may serve as potential biomarkers for guiding treatment decisions, particularly in immuno-oncology, as well as their potential for uncovering novel regulatory mechanisms or therapeutic targets.

\subsection{Strengths and Limitations}

\subsubsection{Strengths}

\paragraph{Integration of Transcriptomics and Clinical Data}
The metagenes were developed from transcriptomic data across a wide gene space. Using variance reduction and ANOVA before clustering ensured the clusters contained genes reflecting differences in immune subtypes. Using clinical data such as survival and cancer type allowed us to evaluate the prognostic value of the metagenes alongside their predictive power for immune subtype. This helps identify which metagenes possess both biological and prognostic relevance, potentially holding promise as biomarkers.

\paragraph{Biologically Interpretable Metagenes}
Through gene-set enrichment analysis using g:Profiler, numerous metagenes were shown to be involved in diverse, immune-related processes in tumorigenesis, like lymphocyte activation, cell-cycle regulation, and ECM remodeling. Further gene-level analysis in section 5.1 uncovered how individual genes functionally complemented one another within the enriched pathways, reinforcing the biological coherence of each metagene. The roles of multiple genes in tumorigenesis and prognosis have been validated by prior studies.

\paragraph{Transparency through SHAP}
Unlike traditional black-box models, SHAP enabled us to quantify the contribution of each metagene-derived feature to model predictions per immune subtype, enabling identification of the transcriptional modules driving each subtype. This improved trust in model predictions, while also facilitating downstream enrichment analysis.

\subsubsection{Limitations}

\paragraph{Lack of External Validation}
The metagenes were derived and evaluated for predictive and prognostic power using patient samples from TCGA. Though this is a large pan-cancer cohort, the metagenes lack external validation using new testing data. Future steps should validate immune state predictions and prognostic value on other cohorts.

\paragraph{Derived Immune Subtype Labels}
It is important to note that this study relies on the work of Thorsson et al. \cite{thorsson2019}. The top 4,489 genes by adjusted p-value from ANOVA were selected based on expression differences between their derived immune subtypes, and the classification models used the metagenes to predict these subtypes. These are pan-cancer subtypes, which means they might miss the immune context based on cancer type. Furthermore, using an existing subtype classification may restrict novel discoveries.

\subsection{Future Directions}

\paragraph{External Validation}
The utmost priority is to validate the identified immune metagenes and the predictive model on independent, external datasets to confirm their generalizability and clinical applicability. The metagenes' prognostic power may be validated with external datasets of transcriptomic and survival data. Metagene predictiveness and prognostic value should also be evaluated per cancer type to see how their role as a biomarker differs across cancer types. Further experimental validation of key metagenes and their constituent genes is crucial to elucidate their precise biological mechanisms in modulating anti-tumor immunity and patient outcomes.

\paragraph{Mechanistic Validation}
Further experimental validation of key metagenes and their constituent genes is crucial to elucidate their precise biological mechanisms in modulating anti-tumor immunity and patient outcomes. In vitro or in vivo validation is required to establish a causal relationship between the metagenes and survival outcomes, possibly through knockdown or overexpression studies.

\paragraph{Therapeutic Implications}
The metagenes identified in the study, particularly those enriched for interferon signaling, immune regulation, and ECM remodeling, may serve as biomarkers of response or resistance to immunotherapies such as checkpoint inhibitors. As part of mechanistic validation, future work should evaluate the clinical impact of overexpression or underexpression of these metagenes in immunotherapy-treated cohorts or those treated with targeted therapies modulating the tumor microenvironment.

\paragraph{Single-Cell Resolution}
Bulk RNA-seq data does not explain which cells are highly expressing the metagenes, or the distribution of metagene expression across the tumor. For this task, single-cell RNA-seq data should be used. Applying similar methodologies to single-cell RNA sequencing data could provide a more detailed understanding of cellular heterogeneity within these immune subtypes. Doing so would also help identify the contribution of specific cell types to a metagene's activity and pinpoint which cellular populations to target for therapeutic intervention.

\section{Conclusion}

This study provided a comprehensive, data-driven, metagene-based framework for characterizing the complex tumor immune landscape. We successfully identified 48 distinct transcriptional metagenes through unsupervised clustering of co-expressed genes. These metagenes represented well-known tumor processes such as ECM remodeling, immune activation and signaling, and cell cycle regulation. Additionally, multiple metagenes showed statistically significant associations with overall survival outcomes, even after adjusting for clinical confounders.

These metagenes were both predictive of tumor immune subtypes and biologically interpretable. While acknowledging the need for validation, both through independent datasets and experimentally, these findings lay the groundwork for developing novel diagnostic, prognostic, and therapeutic strategies tailored to individual immune states in cancer patients.

\end{document}